\documentclass[reprint, 
nofootinbib,
amsmath,amssymb,
aps,prd
]{revtex4}

\usepackage{graphicx}
\usepackage{bm}
\usepackage{booktabs}
\usepackage{amsmath}
\usepackage{mathrsfs}
\usepackage{subfigure}
\usepackage{caption}
\usepackage[hidelinks]{hyperref}
\hypersetup{
    colorlinks=true,
    linkcolor=blue,
    filecolor=gray,      
    urlcolor=blue,
    citecolor=blue,
}
\allowdisplaybreaks[1]

\begin{document}


\title{Measuring Cosmological Redshift Using Gravitational Waves from Compact Binaries with Mass Transfer}

\author{Zi-Han Zhang$^{1,2}$}
 \email{zhangzihan242@mails.ucas.ac.cn}

\author{Tan Liu$^{3,4}$}
 \email{lewton@mail.ustc.edu.cn}

\author{Shenghua Yu$^{5}$}
 \email{shenghuayu@bao.ac.cn}
 
\author{Zong-Kuan Guo$^{3,4,6}$}
\email{guozk@itp.ac.cn}

\affiliation{$^{1}$International Centre for Theoretical Physics Asia-Pacific, University of Chinese Academy of Sciences, 100190 Beijing, China}
\affiliation{$^{2}$Taiji Laboratory for GW Universe, University of Chinese Academy of Sciences, 100049 Beijing, China}

\affiliation{$^{3}$School of Fundamental Physics and Mathematical Sciences, Hangzhou Institute for Advanced Study, University of Chinese Academy of Sciences, Hangzhou 310024, China}

\affiliation{$^{4}$School of Physical Sciences, University of Chinese Academy of Sciences, Beijing 100049, China }

\affiliation{$^{5}$CAS Key Laboratory of FAST, National Astronomical Observatories, Chinese Academy of Sciences, 20A Datun Road, Beijing 100101, China}

\address{$^{6}$Institute of Theoretical Physics, Chinese Academy of Sciences, Beijing 100190, China }


\date{\today}

 \begin{abstract}
The mass transfer process is prevalent during the inspiral phase of compact binary systems. Detection of gravitational waves from the inspiral phase of binaries with white dwarfs will allow us to measure the mass transfer rate. Mass transfer effects provide additional contributions to the phase of gravitational waves, which can break the degeneracy between binary masses and redshift. Based on the analytic mass transfer rate to the first order post-Newtonian evolution of orbital angular frequency, we use the Fisher matrix to forecast the ability of DECIGO to measure the redshift of compact binaries with mass transfer. We conclude that for compact binary systems containing white dwarfs, the redshift can be determined to an accuracy of $10\%$ for $z=0.01$ with a $SNR\thicksim 30$.


\end{abstract}

\maketitle
 
\section{Introduction}
Cosmic expansion can redshift gravitational waves (GWs)\cite{Maggiore:2007ulw,Gravity_PoissonWill,thorneModernClassicalPhysics2018}. Observing GWs can extract the evolution information of the universe and measure the cosmological parameters\cite{Wang_2020}. The  mass $\mathcal{M}$ and the redshift $z$ usually appear as the combination redshifted mass $\mathcal{M}(1+z)$ in gravitational waveforms \cite{CosmologywithGravitational,PhysRevLett.108.091101}.It has been found that the measurement of mass-dependent dimensionless parameters in gravitational waveforms (e.g., mass transfer (MT) rate, tidal deformation) can break the degeneracy of mass-redshift and get the parameter estimation of the redshift\cite{PhysRevLett.108.091101,wg63-t8bp,PhysRevD.85.023535,10.1093/mnrasl/slaa183}.

As the second most numerous population of compact binary systems, neutron star-white dwarf (NSWD) binaries are predicted by LISA to detect 100 to 300 systems form the total number of such systems within the Galaxy may reach $\mathcal{O}(10^6\thicksim10^7)$\cite{Amaro-Seoane2023,Chen_2020}. NSWD systems always involving MT process \cite{10.1093/mnras/stab626,yu2025diversemorphologygravitationalwave,PhysRevD.109.123013}, emitting GWs with frequencies ranging from $0.001$ Hz to $0.8$ Hz with MT rates spanning from $10^{-8}$ to 1 $M_\odot/\text{yr}$ \cite{Pan2023}. These low-frequency GWs within the Galaxy fall within the sensitive frequency bands of space-based GW detectors LISA\cite{NELEMANS201087,Amaro-Seoane2023}, Taiji \cite{10.1093/nsr/nwx116,doi:10.1142/S0217751X2050075X}, and Tianqin \cite{PhysRevD.100.043003}. Beyond the Galaxy, for binary systems located at greater luminosity distances and with larger cosmological redshifts, the Deci-hertz Interferometer Gravitational wave Observatory (DECIGO)\cite{Kawamura_2006,PhysRevLett.87.221103,kawamura2020currentstatusspacegravitational} is expected to detect GW signals from NSWD binaries at a signal-to-noise ratio (SNR) greater than 10 \cite{Burdge2019,10.1093/mnras/sty1545}, while enabling the measurement of redshifts.


The main purpose of this study is to demonstrate a method to extract the redshift $z$ only using GW observation.
Through observing gravitational waves, one can measure the redshifted WD mass $m_{WD}(1+z)$, the redshifted orbital frequency $\omega/(1+z)$ and the MT rate $\chi$. We obtain an analytic expression of the MT rate $\chi$  that depends on the intrinsic WD mass $m_{WD}$ and the intrinsic orbital frequency $\omega$. Using these relations, the cosmological redshift $z$ of the binary system can be extracted.
In the systems with high GW frequency and WD mass, there would be existing a large MT rate about $\chi=10^{-4}M_\odot/\text{yr}$. It is necessary to consider the variation of MT rate and estimate their contribution to the evolution of the orbital angular frequency in Eq.\eqref{variationoemga}. We solve the differential equation Eq.\eqref{variationoemga} by expanding of orbital angular frequency to the order of $\mathcal{O}(t^4)$. We perform a Fisher analysis of GW signals based on DECIGO, and find that the measurement precision of redshift is $\Delta z/z<0.1$ at $z=0.01$, which verifies the detectability of redshift using this method.

This paper is organized as follows. In Sec.\ref{MTrate}, We drive the MT rate as a function of the WD mass and the orbital angular frequency. In Sec.\ref{Frequency}, we derive the evolution of the orbital angular frequency with MT $\chi$, variation of MT $\dot\chi$ and $\ddot \chi$, and quadric MT $\chi^2$ terms. In Sec.\ref{FisherSec}, we do Fisher analysis and compute the projected parameter uncertainties of the
GW detection with DECIGO. We conclude in Sec. \ref{con}.

We use the cosmological parameters $H_0=67.66$ km/s/Mpc, $\Omega_m=0.311$, and $\Omega_\Lambda=1-\Omega_m$\cite{Planck}. Throughout this work, we adopt the unit $G = c = 1$ and $1 ~M_\odot= 4.92\times 10^{-6} ~\text{s} = 1.48 ~\text{km}$.

\section{The evolution of the GW frequency with MT}

\subsection{The mass transfer rate}\label{MTrate}
We consider MT in a compact binary system consisting of a WD and a neutron star (or a black hole). Due to the significant difference in compactness between two components, when the radius of the WD $R_{WD}$ exceeds its Roche radius $R_{RLOF}$, the outer-layer matter of the WD is transferred to the primary star in the form of a common envelope. This MT remains stable with an MT rate of $10^{-5}M_\odot/\text{yr}$ over a timescale of 200 years \cite{10.1093/mnras/stab626,yu2025diversemorphologygravitationalwave}, while an unstable mass transfer with an MT rate of $10^{-3}M_\odot/\text{yr}$ over a timescale of 5 years.

When calculating the analytical expression for the accretion rate, the value of the dimensionless accretion rate depends on the intrinsic properties of the binary system, such as the mass $m_{WD}$ and frequency $\omega$. Here, we use the superscript $``*"$ to denote the intrinsic value of the measured quantities without redshift effect. 

Considering the Roche potential of a binary system, MT rate can be expressed as\cite{2023pbse.book.....T,1988A&A...202...93R,Fernandez_2013,10.1093/mnras/stad1862}
\begin{equation}\label{Chi0}
    \chi=-\dot{m}_{WD}=\mathcal{A}m_{WD}\omega\left(\frac{R_{WD}-R_{RLOF}}{R_{WD}}\right)^3,
\end{equation}
where $\mathcal{A}$ is a dimensionless parameter determined by the internal structure of WD, $R_{RLOF}$ is the Roche radius, $q=m_{WD}/m_{p}$ is the mass ratio of the binary, $r$ is the distance between two stars\cite{2023pbse.book.....T}, 
\begin{equation}\label{Rrlof}
    R_{RLOF} =\frac{0.49q^{2/3}}{0.6q^{2/3}+\ln(1+q^{1/3})}r\simeq 0.49 q^{1/3}r.
\end{equation}
$R_{WD}$ is radius of WD determined by the WD mass and components\cite{1972ApJ175417N}.
 \begin{equation}\label{WDR}
    \quad R_{WD} \simeq 6770.48\left(\frac{m_{WD}}{0.7M_\odot}\right)^{-1/3}\bigg[1-\left(\frac{m_{WD}}{M_{CH}}\right)^{4/3}\bigg]^{1/2}\left(\frac{\mu_e}{2}\right)^{-5/3} M_\odot,
\end{equation}
where $M_{CH}=1.44M_\odot$ is the Chandrasekhar limit and $\mu_e=2$ is the mean molecular weight per electron, which is valid for both He and C/O WDs. Substituting Kepler’s Third Law $r^{3}=GM\omega^{-2}$ and Eqs. \eqref{Rrlof} and \eqref{WDR} into Eq.\eqref{Chi0} yields
\begin{equation}
    \chi=\mathcal{A}m_{WD}\omega\bigg\{1-\mathcal{B}\bigg[qM m_{WD}\bigg[1-\left(\frac{m_{WD}}{M_{CH}}\right)^{4/3}\bigg]^{-3/2}\omega^{-2}\bigg]^{1/3}\bigg\}^3,
\end{equation}
where $\mathcal{B}=7.79\times 10^{-5} ~ M_\odot^{-4}$, the total mass $M=m_{WD}+m_p$ and $m_p$ is the mass of primary start. While the mass of WD much smaller than primary star and the Chandrasekhar mass $m_{WD}\ll m_p \thicksim M_{CH}$, We simplify above equation as 
\begin{equation}\label{Chi*}
    \chi=\mathcal{A}m_{WD}\omega\Big(1-\mathcal{B} m^{2/3}_{WD}\omega^{-2/3}\Big)^3.
\end{equation}
We anticipate $\chi>0$, which means the mass flows from the WD to the primary star. So we have a basic initial condition for MT as $\omega>\mathcal{B}^{3/2}m_{WD}$, and  $\omega > 30 \text{mHz}$ for topical WD mass  $m_{WD}\thicksim 0.2 M_\odot$.

\begin{figure}[ht!] 
    \centering
    \begin{subfigure}[$~m_{WD} - f_{max}$]{{\includegraphics[width=0.48\textwidth]{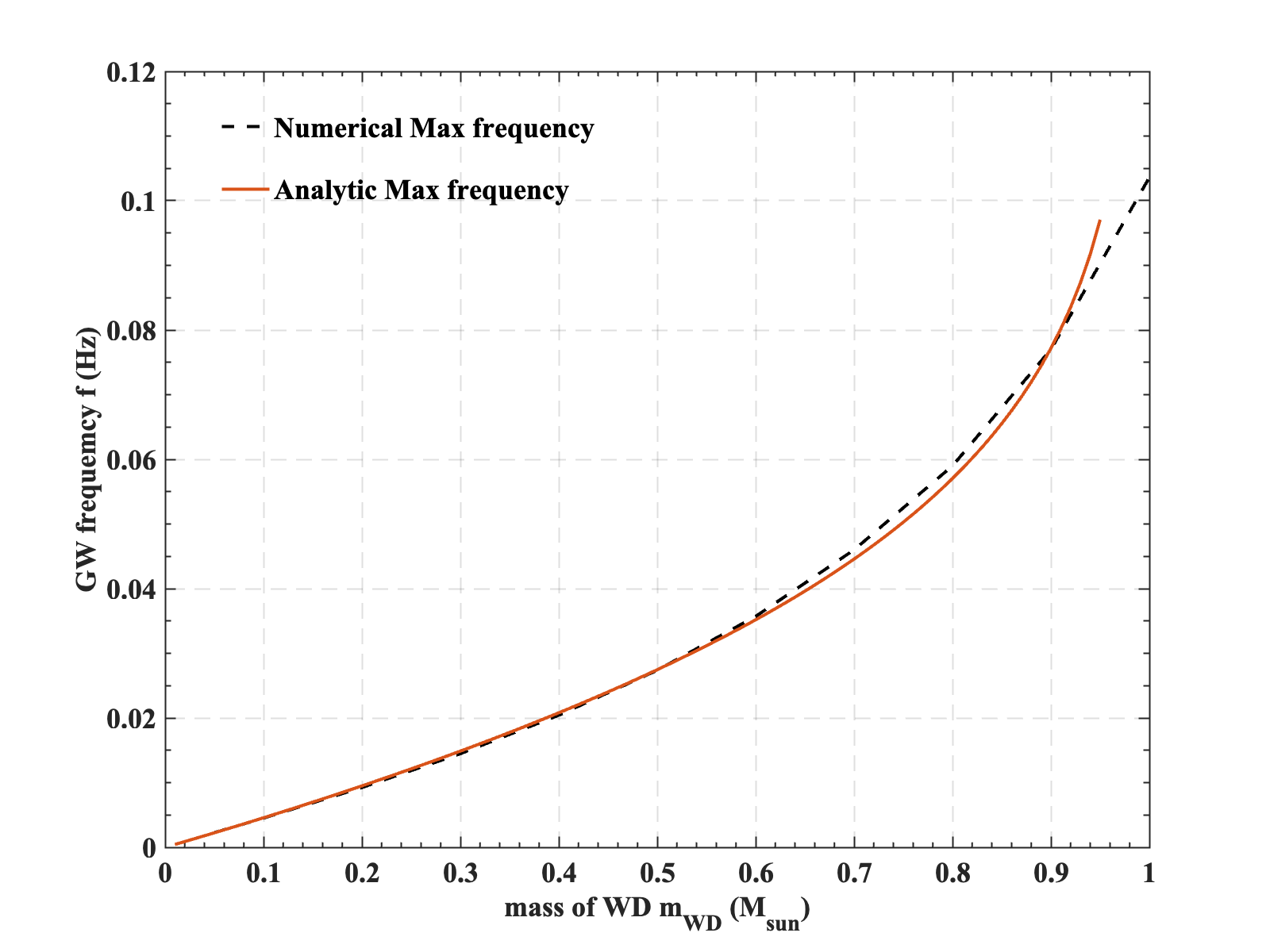}}\label{M_f}}
    \end{subfigure}
    \begin{subfigure}[$~m_{WD} - \chi_{fmax}$]{{\includegraphics[width=0.48\textwidth]{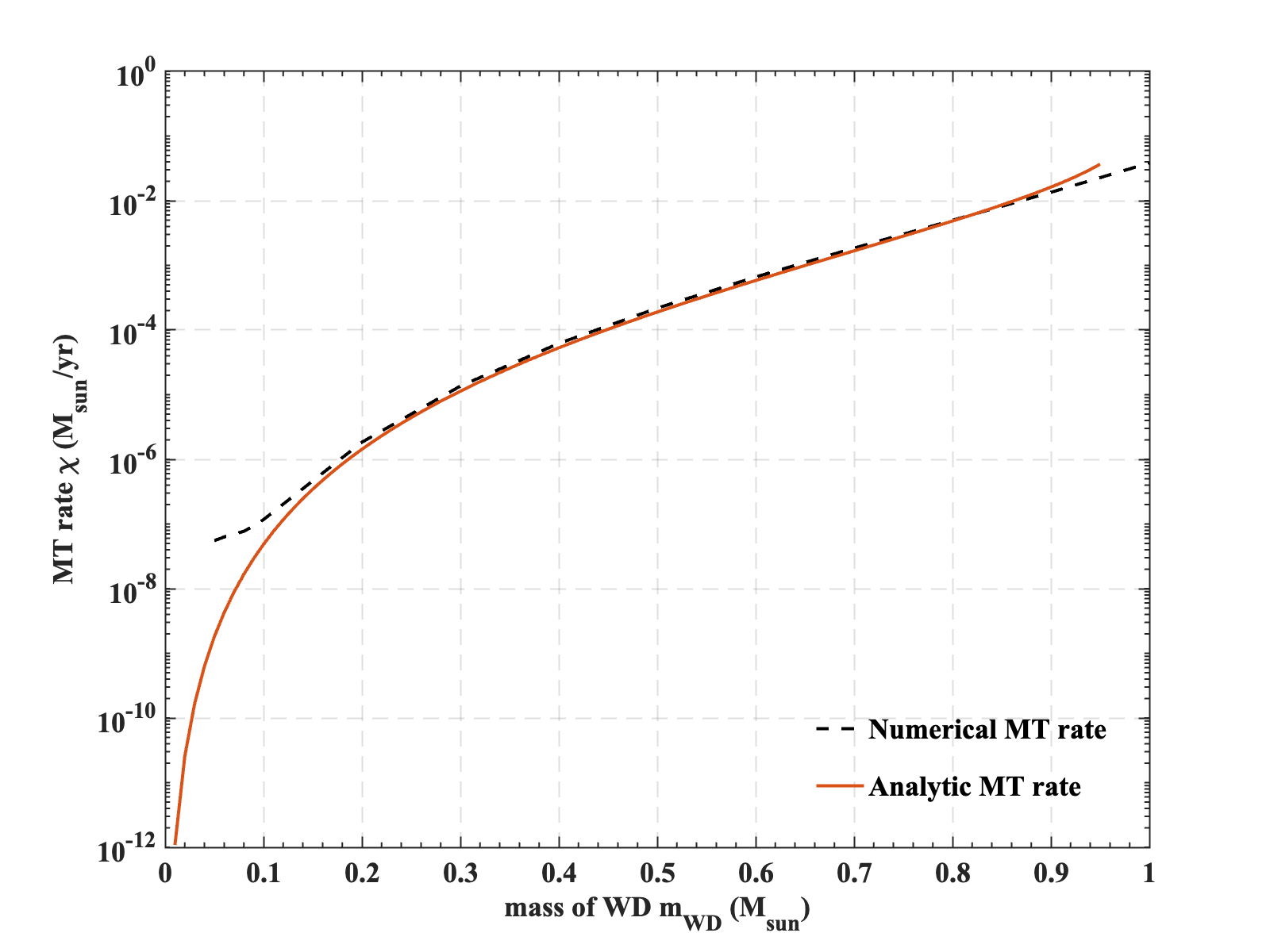}}\label{M_chi}}
    \end{subfigure}
    \caption{(a) Matching between analytic and numerical results of maximum frequency $f_{max}$. The red solid line is the analytic maximum frequency and the black dotted line is the numerical maximum frequency in \cite{10.1093/mnras/stab626}. (b) Matching between analytic and numerical results of MT rate $\chi_{fmax}$ at the maximum frequency. The red solid line is the analytic MT rate obtained by using Eq.\eqref{Chi*} and the black dotted line is the numerical MT rate in \cite{10.1093/mnras/stab626}. When the WD mass ranges from $0.2 M_\odot$ to $0.9 M_\odot$, the analytical results and numerical results exhibit agreement. We have set the value $\mathcal{A}=5\times 10^{-8}$ and $m_p=1.4 M_\odot$.}
    \label{MT_NA}
\end{figure}

It is necessary to verify the correctness and the range of application of the MT rate. Therefore, we matched it with data from numerical simulations  in the Table I of \cite{10.1093/mnras/stab626}, which presents the maximum frequency $f_{max}$ and corresponding MT rate $\chi_{fmax}$ of binary systems with different WD masses. The maximum orbital angular frequency depends on binary masses $m_p$ and $m_{WD}$ and parameter $\mathcal{A}$. We obtain the specific value of the parameter $\mathcal{A}=5\times 10^{-8}$, which matches the numerical results well. Substituting the MT rate in Eq.\eqref{Chi*} into the time derivative of orbital angular frequency in Eq.\eqref{variationoemga}, and set the time derivative the orbital angular frequency $\dot\omega$ to 0  to obtain the maximum orbital angular frequency $\omega_{max}$ as well as the corresponding MT rate $\chi_{fmax}$ at the maximum frequency. By setting the initial masses with $m_p=1.4 M_\odot$ and $m_{WD}$ ranging from $0.01 M_\odot$ to $1 M_\odot$, the analytical relationship between $m_{WD}$ and $f_{max}$ is shown as the red solid line in Fig.\ref{M_f}, where $f_{max}=\omega_{max}/\pi$ is the maximum GW frequency. Substituting the maximum frequency into the MT rate in Eq.\eqref{Chi*} yields the analytical relationship between $m_{WD}$ and $\chi_{fmax}$, which is shown as the red solid line in Fig.\ref{M_chi}.

In the two subfigures of Fig.\ref{MT_NA}, it can be seen that when the WD mass ranges from $0.2 M_\odot$ to $0.9 M_\odot$, the analytical results and numerical results for both the maximum frequency and the corresponding MT rate match well while choosing the value $\mathcal{A}=5\times 10^{-8}$. Therefore, in the following calculations, the WD mass will be constrained within the range of $0.2 M_\odot$ to $0.9 M_\odot$.

 \begin{figure}[ht!]
    \centering
    {\includegraphics[width=0.48\textwidth]{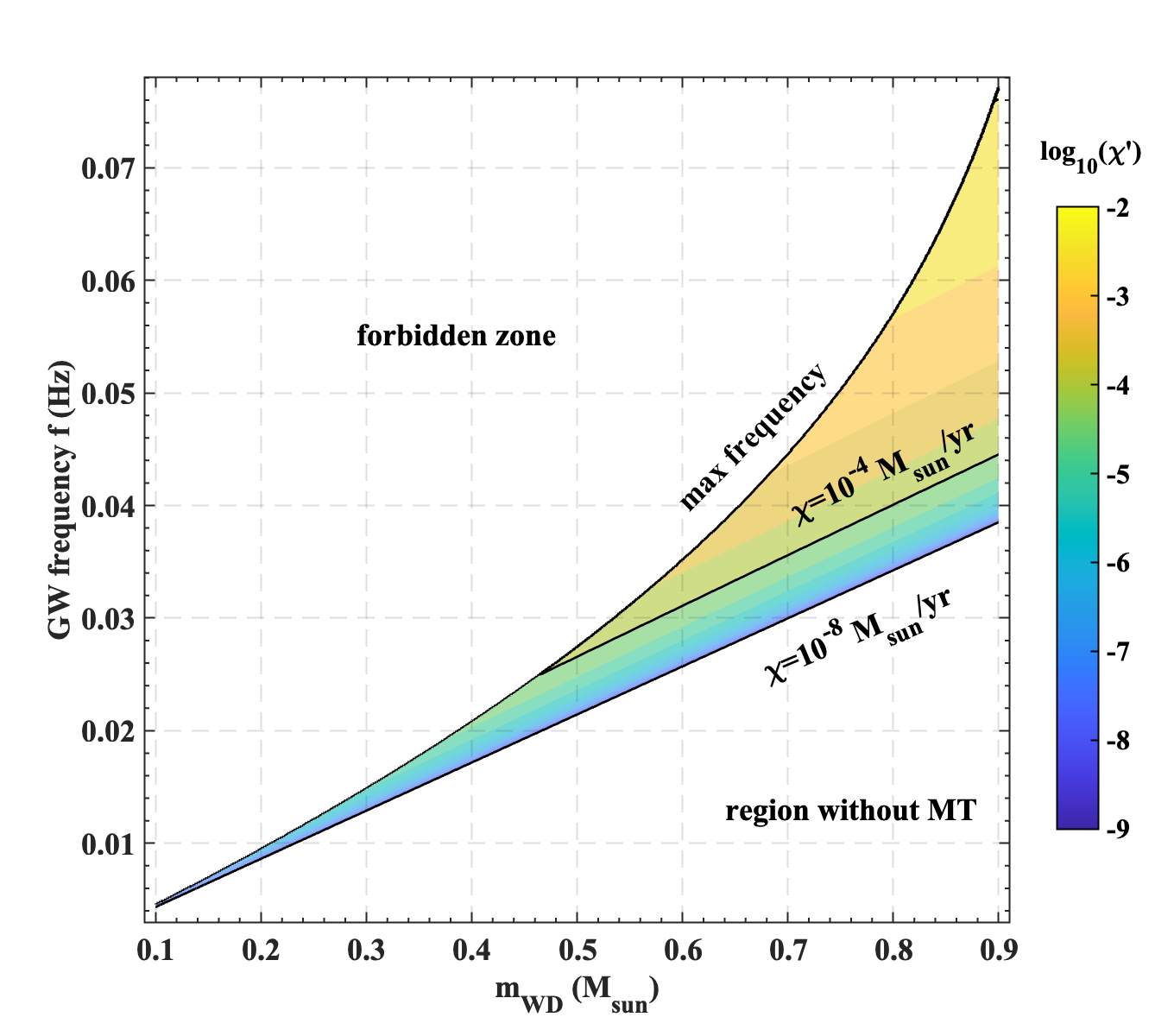}}
    \caption{The relationship between MT rate $\chi$, GW frequency $f$, and WD mass $m_{WD}$ with the mass of primary star $m_p=1.4 M_\odot$.  $\chi'=\chi/(1 M_\odot/\text{yr})$ is the dimensionless MT rate. The ``forbidden zone" refers to the frequency range that cannot be reached for a given WD mass or the frequency higher than the maximum frequency. The ``region without MT" denotes the area where MT has not yet occurred because the WD radius is smaller than the Roche lobe radius. The MT rate increases with the increase of frequency while with the decrease of WD mass.}
    \label{MTRate_m_f}
\end{figure}

In FIG.\ref{MTRate_m_f}, we plot the influence of two independent variables, GW frequency $f$ and WD mass $m_{WD}$, on the MT rate $\chi$. The ``forbidden zone" refers to the frequency range that cannot be reached for a given WD mass or the frequency higher than the maximum frequency. The ``region without MT" denotes the area where MT has not yet occurred because the WD radius is smaller than the Roche lobe radius. In our calculations, the valid parameter space for frequency and white dwarf mass corresponds to the colored region in FIG.\ref{MTRate_m_f}. When the MT rate is less than $10^{-4}M_\odot$, we consider the binary system to be in a stable MT process, the MT variation is much smaller than the MT rate itself, and the mass of the formed accretion disk is negligible \cite{Röpke2023,Ivanova2013}. Eq.\eqref{variationoemga} is more accurate for binary systems undergoing stable MT process.

\subsection{The evolution of the GW frequency and the phase}\label{Frequency}

The quadrupole GW waveforms of the inspiral binary with MT is given by \cite{Maggiore:2007ulw}
\begin{align}
    h_+(t)&=\frac{4}{D}\left(\frac{G\mathcal{M}_c(t)}{c^2}\right)^{5/3}\left(\frac{\pi f(t)}{c}\right)^{2/3}\frac{1+\cos^2\iota }{2}\cos\psi(t),\label{ht}\\
    h_\times(t)&=\frac{4}{D}\left(\frac{G\mathcal{M}_c(t)}{c^2}\right)^{5/3}\left(\frac{\pi f(t)}{c}\right)^{2/3}\cos\iota\sin\psi(t).\label{ht2}
\end{align}
where $\mathcal{M}_c(t)=(m_{WD}(t)m_p(t))^{3/5}/M^{1/5}$ is the chirp mass and $D$ is the  distance from the source to our detectors. $\iota$ is the angle between the line of sight and the angular momentum of the binary system. The GW phase can also be easily expressed in terms of the effects of GW radiation and MT. The phase caused by MT effects $\psi^{MT}$ and $\psi^{GW-MT}$ is the key to break the degeneracy between mass and redshift.
\begin{equation}
    \psi(t)=\psi_0+\psi^{GW}(t)+\psi^{MT}(t)+\psi^{GW-MT}(t),
\end{equation}
where $\psi_0$ is the initial GW phase. The frequency of GW is $f=\omega/\pi$. The phase of GWs can be calculated by integrating of the angular frequency $2\pi f(t)$ as $\psi(t)=2\pi\int_0^t f(t')\mathrm{d}t'$, then we get
\begin{equation}
        \psi(t)^A=2\pi \left[f_0^A t+\sum_{i=1}^3 \frac{1}{(i+1)!}f_{i}^A t^{i+1}+\mathcal{O}\left(t^5\right)\right],\label{psiL}
\end{equation}
where $f^A_i=\omega^A_i/\pi$ is the expansion terms of $f$ and $i=0,1,2,3$ in Eq.\eqref{omega}.  Superscripts ``A" can be replace by ``GW", ``MT", and ``GW-MT". The superscript ``GW" denotes the frequency evolution caused by GW radiation, ``MT" denotes the frequency evolution caused by GW effects, and ``GW-MT" denotes the contribution of GW radiation couples with MT effects.

Based on the calculation results from our previous work \cite{qhf1-v5jr}, the time derivative of the orbit frequency $\omega$ in quasi-circular orbits under the combined influence of gravitational radiation and MT effects, with the precision retained up to the 1PN order, is given by
\begin{align}\label{variationoemga}
    \frac{\dot{\omega}}{\omega}=&-3\bigg\{1+\frac{1}{c^2}\bigg(3+\frac{1}{2}\eta\bigg)(GM)^{2/3}\omega^{2/3}\bigg\}\Delta\left(\frac{\chi}{M\eta}\right)-\frac{1}{c^2}\frac{3}{2} G M \eta \left(\frac{\chi}{M\eta}\right)^2\\\nonumber
    &+\frac{1}{c^5}\bigg\{1+\frac{1}{c^2}\bigg(-\frac{743}{336}-\frac{11}{4}\eta\bigg)(GM)^{2/3}\omega^{2/3}\bigg\}\mathcal{F}\eta\omega^{8/3},
\end{align}
where $\mathcal{F}=(96/5)(GM)^{5/3}$, $\eta(t)=m_pm_{WD}/M^2$ is the dimensionless symmetry mass ratio, and $\Delta(t)=(m_p-m_{WD})/M$ is the dimensionless mass difference. The terms proportional to $\mathcal{F}$ originate from the gravitational radiation backreaction and the terms depending on $\chi/(M\eta)$ originate from the MT effect. The MT rate $\chi$ varies with time here, whereas in most previous works, the MT was treated as a constant \cite{10.1093/mnras/stad2358}.
\begin{equation}\label{Deltat}
    \Delta(t)=\Delta_0+2\left(\frac{\chi_0}{M}\right)t+\left(\frac{\dot\chi_0}{M}\right)t^2+\frac{1}{3}\left(\frac{\ddot\chi_0}{M}\right)t^3+\mathcal{O}(t^4),
\end{equation}
\begin{equation}\label{Etat}
    \eta(t)=\eta_0-\Delta_0 \left(\frac{\chi_0}{M}\right)t-\left[3 \left(\frac{\chi_0}{M}\right)^2+ \frac{1}{2}\Delta_0\left(\frac{\dot\chi_0}{M}\right)\right]t^2-\left[3 \left(\frac{\chi_0}{M}\right)\left(\frac{\dot\chi_0}{M}\right)+ \frac{1}{6}\Delta_0\left(\frac{\ddot\chi_0}{M}\right)\right]t^3+\mathcal{O}(t^4).
\end{equation}
The detail analysis about MT variation terms and quadratic MT terms is in Sec.\ref{Redshift} and Appendix.\ref{App.A}. We get the time evolution of the orbital frequency by the Taylor expansion at $\omega_0$
\begin{equation}\label{omega}
    \omega=\omega_0+\omega_1 t+\frac{1}{2!}\omega_2 t^2+\frac{1}{3!}\omega_3 t^3+\mathcal{O}(t^4),
\end{equation}
where $\omega_1=\dot{\omega}|_{\omega_0}, \omega_2=\ddot{\omega}|_{\omega_0}, \omega_3=\dddot{\omega}|_{\omega_0}$, and $\omega_1=\omega_1^{GW}+\omega_1^{MT}$. 
\begin{align}
    \omega_1^{GW}=& \mathcal F\eta_0 \left[1 + \frac{1}{c^{2}} \left(GM\right)^{2/3} \left(-\frac{743}{336} - \frac{11}{4}\eta_0\right) \omega_0^{2/3}\right] \omega_0^{11/3},\\
    \omega_1^{MT}=&-3\left(\frac{\chi_0 }{M \eta_0}\right)\Delta_0\left[1 + \frac{1}{c^{2}} \left(GM\right)^{2/3} \left(3 + \frac{1}{2}\eta_0\right) \omega_0^{2/3}\right] \omega_0,
    \end{align}
and $\omega_2=\omega_2^{GW}+\omega_2^{MT}+\omega_2^{GW-MT}$, where 
\begin{align}
    \omega_2^{GW}=&\mathcal{F}^2\eta_0^2\left[\frac{11}{3} +\frac{1}{c^2} (G M)^{2/3}\omega_0^{2/3} \left(-\frac{743}{42}-22 \eta_0\right) \right]\omega_0 ^{19/3}\\
    \omega_2^{MT}=&\left(\frac{\chi_0 }{M \eta_0}\right) ^2 \left[6-30 \eta_0 +\frac{1}{c^2} (G M)^{2/3}\omega_0^{2/3} \bigg(63-258 \eta_0-51 \eta_0 ^2 \bigg)  \right]\omega_0\nonumber\\
   +&\left(\frac{\dot\chi_0 }{M \eta_0}\right)\Delta_0\left[-3-\frac{1}{c^{2}} (G M)^{2/3} \omega_0^{2/3}\left(9+\frac{3}{2}\eta_0 \right) \right]\omega_0,\\
   \omega_2^{GW-MT}=&\left(\frac{\chi_0 }{M \eta_0}\right) \mathcal{F} \Delta_0\eta_0\left[-15 +\frac{1}{c^2} (G M)^{2/3}\omega_0 ^{2/3} \left( -\frac{3497}{336} +\frac{83}{2} \eta_0 \right) \right]\omega_0 ^{11/3},
   \end{align}
and $\omega_3=\omega_3^{GW}+\omega_3^{MT}+\omega_3^{GW-MT}$, where 
\begin{align}
   \omega_3^{GW}=&\mathcal{F}^3 \eta_0 ^3\left[\frac{209}{9} +\frac{1}{c^2} (G M)^{2/3}\omega_0 ^{2/3} \left(-\frac{529759}{3024}-\frac{7843}{36}  \eta_0\right) \right]\omega_0 ^9,\\
   \omega_3^{MT}=&\left(\frac{\chi_0 }{M \eta_0}\right) ^3 \Delta_0\left[ -6 +60 \eta_0+\frac{1}{c^2} (G M)^{2/3} \omega_0 ^{2/3} \bigg(-243+1293 \eta_0  +300 \eta_0 ^2 \bigg)\right]\omega_0\nonumber\\
   +&\left(\frac{\chi_0 }{M \eta_0}\right)\left(\frac{\dot\chi_0 }{M \eta_0}\right) \left[18-90\eta_0+\frac{1}{c^2} (G M)^{2/3} \omega_0^{2/3} \bigg(189-774\eta_0-153\eta_0^2\bigg)\right]\omega_0\nonumber\\
   +&\left(\frac{\ddot\chi_0 }{M \eta_0}\right)\Delta_0\left[-3+\frac{1}{c^2} (G M)^{2/3} \omega_0^{2/3} \left(-9 -\frac{3}{2}\eta_0\right)\right]\omega_0,\\
    \omega_3^{GW-MT}=&\left(\frac{\chi_0 }{M \eta_0}\right) \Delta_0 \mathcal{F}^2\eta_0^2 \left[-132 +\frac{1}{c^2} (G M)^{2/3} \omega_0 ^{2/3} \left(\frac{19757}{72}+\frac{3297}{4} \eta_0  \right)\right]\omega_0 ^{19/3}\nonumber\\
   +&\left(\frac{\chi_0 }{M \eta_0}\right) ^2 \mathcal{F}\eta_0 \left[171 -720 \eta_0  +\frac{1}{c^2} (G M)^{2/3}\omega_0 ^{2/3} \left(\frac{242603 }{336}-\frac{95779 }{28}\eta_0+\frac{4149 }{2}\eta_0 ^2\right)\right]\omega_0 ^{11/3}\nonumber\\
   +&\left(\frac{\dot\chi_0 }{M \eta_0}\right)\Delta_0\mathcal{F}\eta_0\left[-18  +\frac{1}{c^2} (G M)^{2/3} \omega_0^{2/3} \left(-\frac{1577}{84}+\frac{189}{4 } \eta_0\right)\right]\omega_0^{11/3}.
\end{align}
Terms with $(\chi/M)$ denote the order of MT, terms with $(\dot\chi/M)$ denote the order of MT variation, and terms with $\mathcal{F}$ denote the order of GW, and terms with $c^{-2}$ denote the PN order.

Compared with previous works \cite{PhysRevD.111.043049}, the above expansion method for frequency is no longer limited to the ``low-frequency" band of GWs, which improves the calculation accuracy for high frequencies. 

 \begin{figure}[ht!] 
    \centering
    {\includegraphics[width=0.5\textwidth]{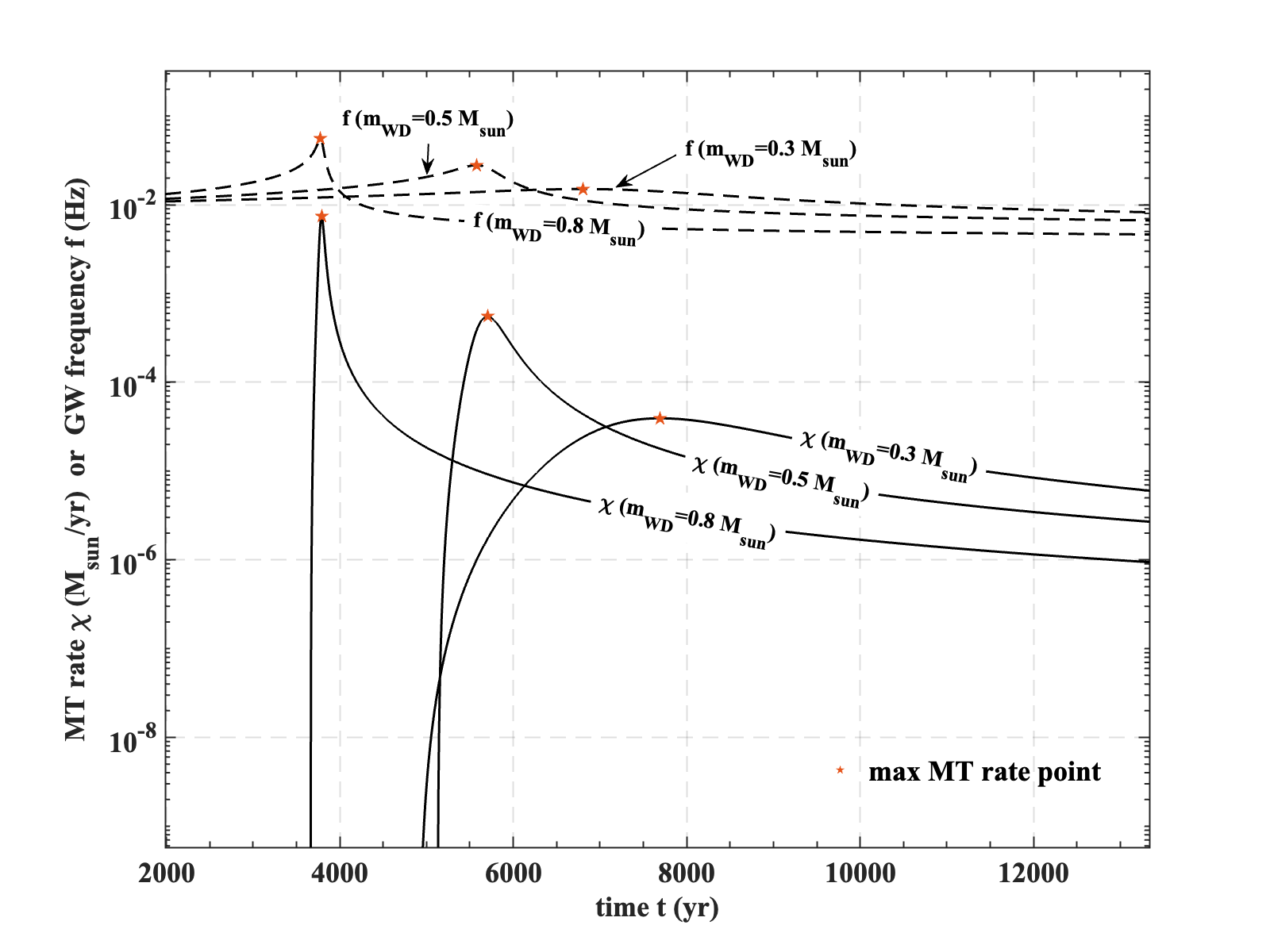}}
    \caption{The evolution of MT rate and GW frequency for different initial WD masses. The dashed lines represent the evolution of GW frequency, the solid lines represent the evolution of MT rate, and the red stars denote the maximum points of each curve. The initial conditions are $f_0=0.01$ Hz, $m_p=0.2 M_\odot$, and $m_{WD}=0.3,0.5,0.8 M_\odot$.
}
    \label{Mt}
\end{figure}

Fig. \ref{Mt} shows the MT rates calculated by Eq.\eqref{Chi*} and GW frequency for different WD masses. We initiate the evolution of the binary system from a low GW frequency, during which no MT occurs. After approximately 4000 years of evolution, the WD with a mass of $0.8M_\odot$ begins to undergo MT. When the MT rate is small, the frequency evolution is dominated by gravitational radiation, the GW frequency exhibits a chirping trend of increase, while the MT rate rises simultaneously. Once the effect of MT exceeds that of gravitational radiation, the GW frequency reaches its maximum value and then continues to decrease thereafter. The maximum point of the MT rate occurs later than that of the GW frequency, which is clearly evident from the curve corresponding to a WD mass of $m_{WD}=0.3 M_\odot$ in Fig. \ref{Mt}.

Owing to the properties of WDs, it can be inferred from Eq. \eqref{WDR} that the radius of a WD increases as its mass decreases. Therefore, even though the binary is gradually moving apart, the WD's radius remains larger than the Roche radius, ensuring that the MT process persists continuously. Without external forces, the binary will keep moving farther apart and tend to a constant orbital frequency \cite{10.1093/mnrasl/slx064}.

\subsection{Cosmological redshift effect}\label{Redshift}
The cosmological redshift effect, not incorporated into the gravitational waveforms in the previous subsection, should be included through the following procedures \cite{wg63-t8bp,thorneModernClassicalPhysics2018}. The source distance $D$  should be replaced by the luminosity distance $\mathcal{D}$,  the mass parameter $m$ should be replaced by the redshifted mass parameter $m(1+z)$. The  parameter of dimension $[\text{mass}]^n$ should be multiplied by $(1+z)^n$. 
Since the MT rate $\chi$ is dimensionless, it is not influenced by the cosmological redshift effect.


Through observing gravitational waves, one can measure the redshifted WD mass $m_{WD}(1+z)$ and the MT rate $\chi$. Since the MT rate $\chi$ depends on the intrinsic WD mass $m_{WD}$, the cosmological redshift $z$ of the binary system can be extracted.
We use the overhead bar to indicate the redshifted quantities, such as $\bar{m}=m(1+z)$ and $\bar{\omega}=\omega(1+z)^{-1}$. Then, Eq.\eqref{Chi*} can be rewritten as follows:
\begin{equation}\label{Chiredshift}
        \chi=\mathcal{A}\bar{m}_{WD}\bar{\omega}\Big[1-\mathcal{B} \bar{m}^{2/3}_{WD}\bar{\omega}^{-2/3}(1+z)^{-4/3}\Big]^3.
\end{equation}

\section{detectability of the cosmological redshift}\label{FisherSec}

\subsection{Fisher Analysis}
The  Fisher matrix formalism allows us to compute the uncertainties associated with the measurement of a set of signal parameters. The noise curve of DECIGO is \cite{Kawamura_2006, 10.1143/PTP.123.1069, PhysRevD.83.044011} 
\begin{equation}
 S_n(f)=6.53\times 10^{-49}\bigg[1+\left(\frac{f}{7.36\text{ Hz}}\right)^2\bigg]+4.45\times10^{-51}\left(\frac{f}{1\text{ Hz}}\right)^{-4}\bigg[1+\left(\frac{f}{7.36\text{ Hz}}\right)^2\bigg]^{-1}+4.94\times10^{-52}\left(\frac{f}{1\text{ Hz}}\right)^{-4} \text{ Hz}^{-1}.
\end{equation}
The signal-to-noise ratio (SNR) is given by \cite{PhysRevD.49.2658,Takahashi_2002}
\begin{equation}
\begin{split}
    \left(\frac{S}{N}\right)^2&=4\int^\infty_0\frac{|\Tilde{s}(f)|^2}{S_n(f)}\mathrm{d}f\approx\frac{2}{S_n(f_0)}\int^{T_{obs}}_0 |s(t)|^2\mathrm{d}t.
\end{split}
\end{equation}
where $\Tilde{s}(f)$ is the Fourier transforms of $s(t)$ and $T_{obs}$ is the observation time.  When the SNR is large enough, the parameter-estimation errors $\Delta \theta^i$ can be determined by the Fisher matrix \cite{PhysRevD.46.5236}
\begin{equation}\label{Gamma}
    \Gamma_{ij}\equiv\left(\frac{\partial s}{\partial \theta^i}\bigg|\frac{\partial s}{\partial \theta^j}\right).
\end{equation}
The inner product $(s_1|s_2)$ of two signals $s_1(t)$ and $s_2(t)$ is \cite{PhysRevD.49.2658}
\begin{equation}(s_1|s_2)=2\int^\infty_{0}\frac{\Tilde{s}^*_1(f)\Tilde{s}_2(f)+\Tilde{s}_1(f)\Tilde{s}^*_2(f)}{S_n(f)}\mathrm{d}f.
\end{equation}
The root-mean-square error in $\theta^i$ is
\begin{equation}
\sqrt{\left<(\Delta\theta^i)^2\right>}=\sqrt{\Sigma^{ii}},\;\;\;\bm{\Sigma}=\bm{\Gamma}^{-1}.
\end{equation}
We consider 6 independent parameters  $\theta_i=\{\mathcal{D},\psi_{0},\bar{m}_{p0},\bar{m}_{WD0},\bar{f}_0,z\}$. For simplicity, the orientation of the binary system is set to be face-on, with the antenna pattern functions given by $F_+=1$ and $F_\times=0$.  According to above derivations, the partial derivatives of the GW signal with respect to each observable quantity in Eq.\eqref{Gamma} can be calculated using completely analytical expressions to ensure the accuracy of the Fisher analysis.

\subsection{Numerical results}
We consider binary systems with different WD masses. The other parameters of the bianry systems are given by  $\mathcal{D}=40 \text{Mpc}, \psi_0=\pi, \text{and }  \bar{m}_p=1.4 M_\odot$. The results of the Fisher analysis for the five-year DECIGO observation are presented in the table below.
\begin{table}[htbp!]
\small
  \centering
  \caption{The SNR and parameter estimation uncertainties for binary systems with different WD masses and maximum frequencies observed by DECIGO. The precision of the redshift measurement is enhanced with increasing white dwarf mass. When the redshifted  WD mass is  0.6 $M_\odot$, the redshift can be measured with an accuracy about $10\%$ for $z=0.01$.}
    \renewcommand\arraystretch{1.6}
    \setlength{\tabcolsep}{6pt}
    \begin{tabular}{ccccccccc}
    \hline
     $\bar{f}_{max}$ (Hz)&$\bar{m}_{WD}(M_\odot)$&SNR    &  $\chi~(M_\odot/\text{yr})$ & $\Delta\psi_{0}/\psi_{0}$  &  $\Delta \bar{m}_{p}/\bar{m}_{p}$ &  $\Delta \bar{m}_{WD}/\bar{m}_{WD}$ &  $\Delta \bar{f}_0/\bar{f}_0$ &  $\Delta z/z$ \\
    \hline
    0.009&0.2& 0.27&$4.37\times10^{-9}$   & $2.87$ & $1.45\times10^3$ & $3.10\times10^3$   & $0.58$ &$1.66\times10^3$\\
    0.014&0.3& 1.26&$6.52\times10^{-7}$   & $0.59$ & $1.34\times10^2$ & $4.03\times10^2$   & $0.007$ &$4.14\times10^2$\\
    0.020&0.4& 4.22&$1.66\times10^{-5}$   & $0.17$ & $0.26$ & $9.34$   &$1.40\times10^{-3}$& $14.60$ \\
    0.027&0.5& 11.44&$1.18\times10^{-4}$   & $0.06$ & $0.12$ & $0.43$   &$4.92\times10^{-5}$& $1.16$ \\
    0.036&0.6& 28.81&$6.25\times10^{-4}$   & $0.03$ & $0.01$ & $0.01$   &$9.36\times10^{-7}$& $0.13$ \\
    0.046&0.7& 63.03&$1.91\times10^{-3}$   & $0.01$ & $1.37\times10^{-3}$ & $1.94\times10^{-4}$   &$1.17\times10^{-8}$& $0.027$ \\
    0.059&0.8& 135.67&$5.40\times10^{-3}$   & 0.006 & $2.46\times10^{-4}$ & $3.22\times10^{-5}$   &$2.90\times10^{-9}$& $0.006$ \\
    \hline
    \end{tabular}%
  \label{fisher}%
\end{table}%

In Table \ref{fisher}, with the increase of the redshifted contact frequency $\bar{f}_0$ and the redshifted WD mass $\bar{m}_{WD}$, both the MT rate and the SNR of gravitational wave detection by DECIGO increase. When the SNR is above 10, the redshift reaches a detectable precision $\Delta z/z\thicksim 1$.

 \begin{figure}[ht!] 
    \centering
    {\includegraphics[width=0.48\textwidth]{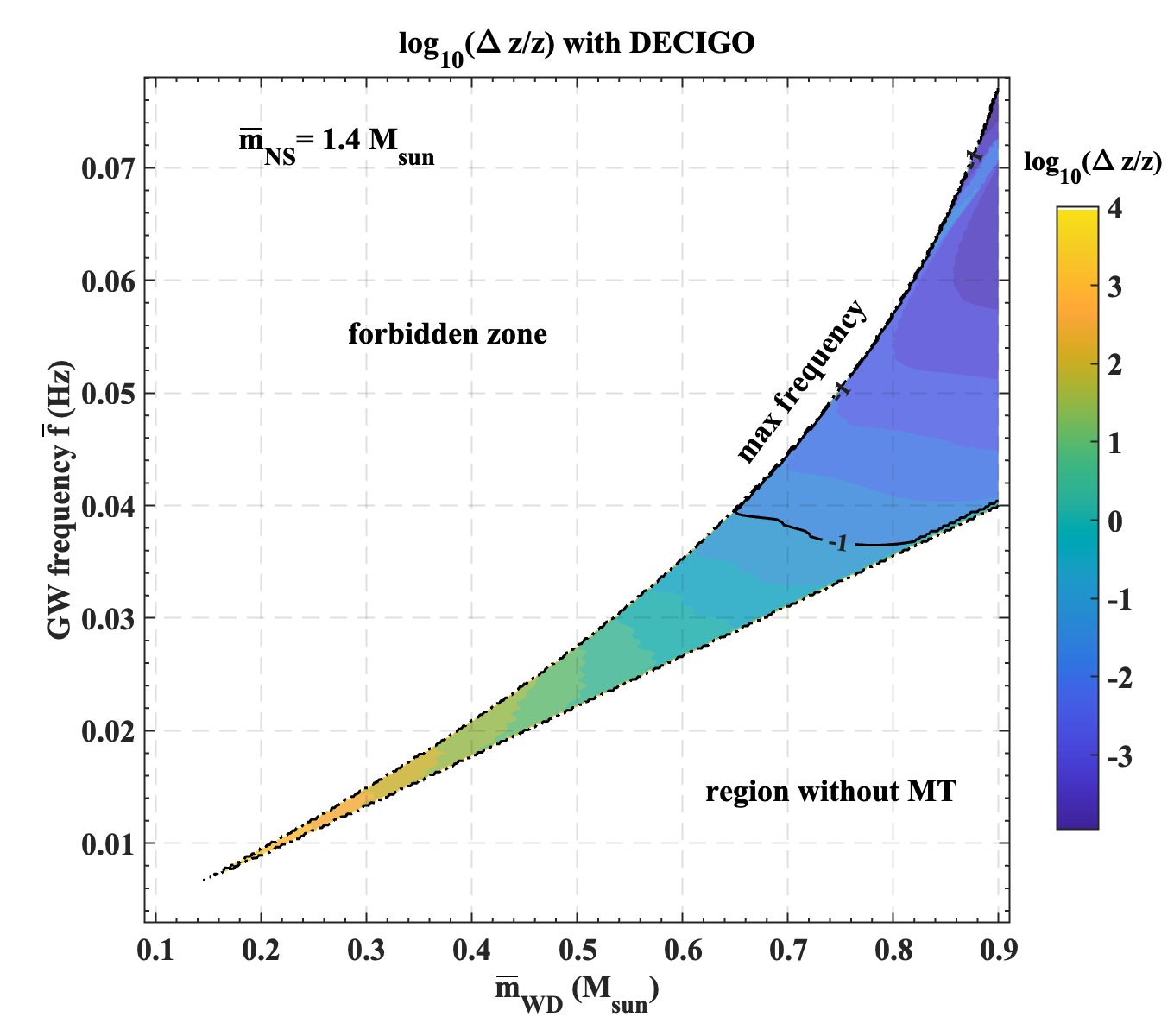}}
    {\includegraphics[width=0.48\textwidth]{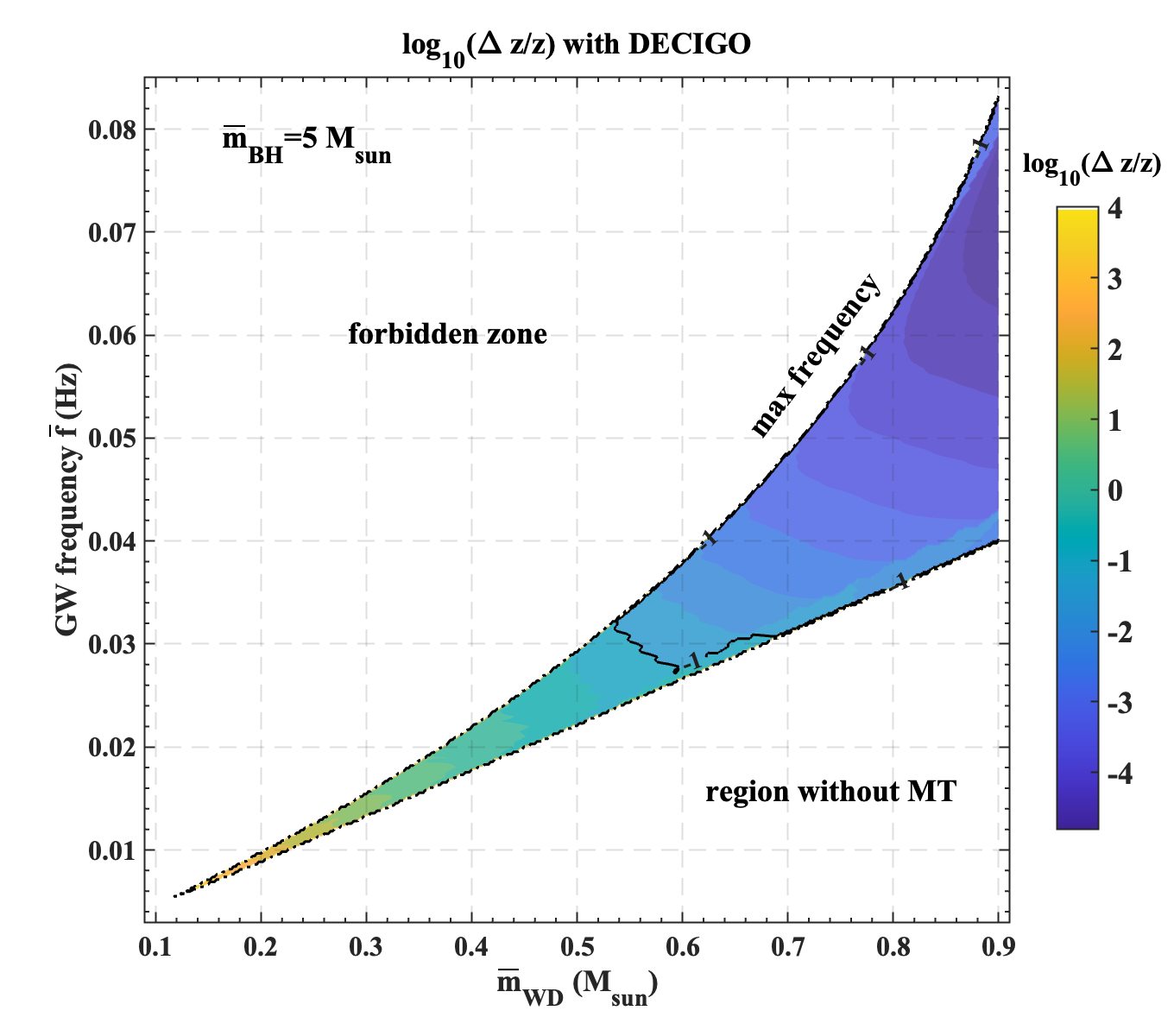}}
    \caption{The relative uncertainty of redshift under different GW frequencies and WD masses. The primary star in the left figure is a $1.4 M_\odot$ neutron star, while the one in the right figure is a $5 M_\odot$ black hole. Within the parameter space of the blue region above the contour line of $\log(\Delta z/z)=-1$, the redshift $z$ can be detected with an accuracy higher than $10\%$.  The luminosity distance of the binary systems is 40 Mpc and the redshift is 0.01.}
    \label{Redshift_mf}
\end{figure}

In Fig. \ref{Redshift_mf}, the redshift uncertainty is presented for the values of frequency and mass that fall within the valid parameter space shown in Fig. \ref{MTRate_m_f}. Taking a precision of $10\%$ as the threshold, the left figure shows that a neutron star - white dwarf binary systems containing WDs with a redshifted mass below 0.65$M_\odot$ are unable to break the mass-redshift degeneracy while the mass of the neutron star is $m_{NS}=1.4 M_\odot$. This is attributed to the relatively low SNR and MT rate in these situations. There would be black hole - white dwarf binary systems as well \cite{10.1093/mnras/stx166,yang2025formationpossibleblackholeultracompact}, which are shown in the right figure in Fig.\ref{Redshift_mf} with a black hole redshifted mass $\bar{m}_{BH}=5 M_\odot$. As the WD mass and frequency increase, the detection precision of redshift improves accordingly. Binary systems with high orbital frequency and high WD mass are capable of breaking the mass-redshift degeneracy. 


 \begin{figure}[ht!] 
    \centering
    {\includegraphics[width=0.48\textwidth]{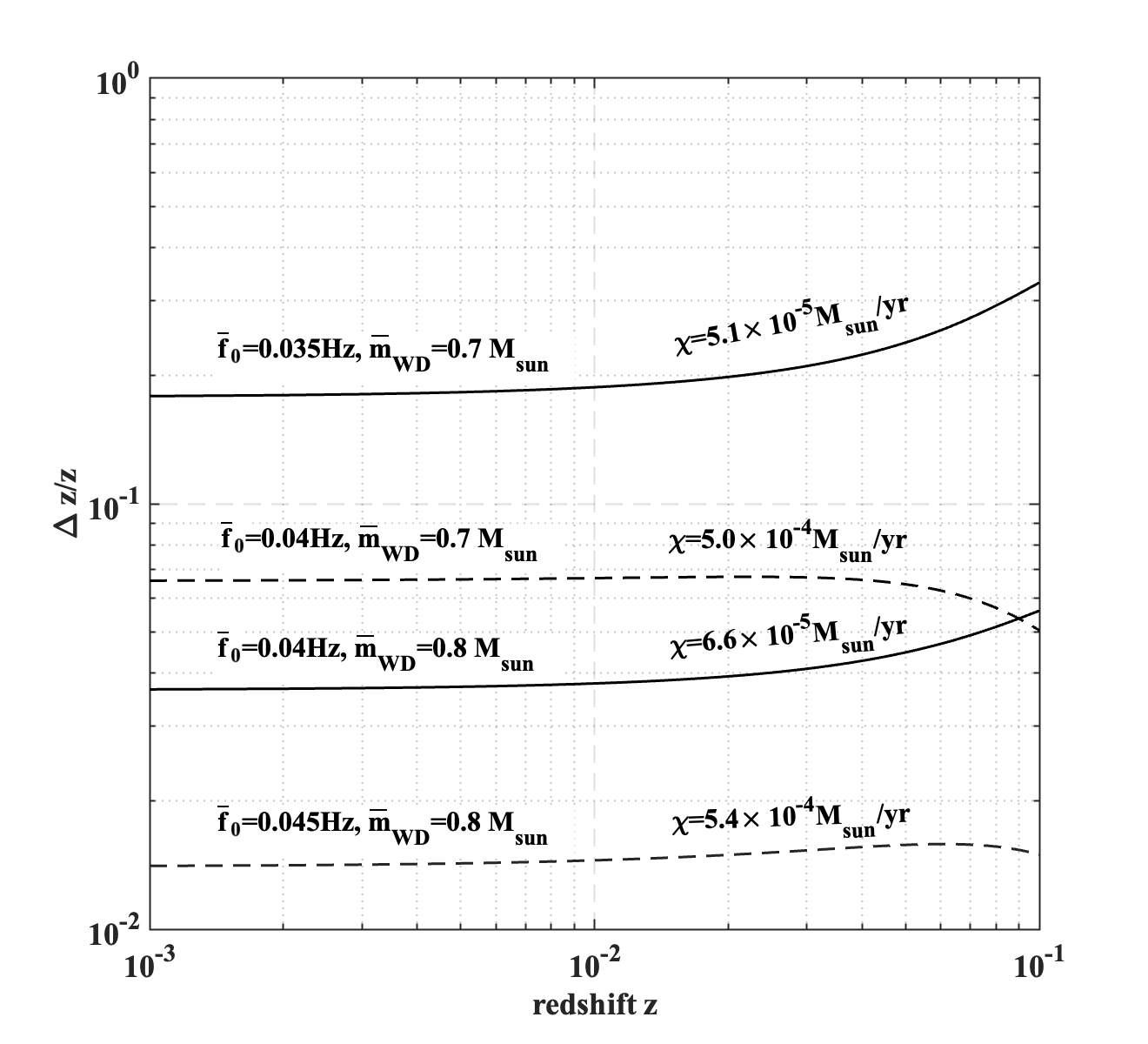}}
    \caption{The relative uncertainty in redshift measurements of binaries across different redshifts. The redshift uncertainty under four sets of parameters $(\bar{f}_0=0.035\text{Hz}$ and $ \bar{m}_{WD}=0.7M_\odot)$, $(\bar{f}_0=0.04\text{Hz}$ and $ \bar{m}_{WD}=0.7M_\odot)$, $(\bar{f}_0=0.04\text{Hz}$ and $  \bar{m}_{WD}=0.8M_\odot)$, and $(\bar{f}_0=0.045\text{Hz}$ and $  \bar{m}_{WD}=0.8M_\odot)$ are plotted as functions of redshift $z$ ranging from $10^{-3}$ to $0.1$.}
    \label{Deltaz}
\end{figure}

Fig. \ref{Deltaz} shows the relative uncertainty in redshift measurements of binaries across different redshifts.
When the binary system reaches $z\thicksim0.1$, the SNR of GWs detected by DECIGO is approximately $1$; even if the redshift uncertainty is small, the signal cannot be detected. When the redshift ranges from $10^{-3}$ to $10^{-2}$, the redshift uncertainty remains stable. Moreover, when the GW frequency is higher than $0.035$Hz or the mass is greater than 0.7$M_\odot$, the redshift can be measured with an uncertainty of less than $18\%$.

\section{Conclusion}\label{con}
We have derived an analytical expression for the MT rate, which depends on the WD mass and the orbital angular frequency of the binary system. Since the MT rate is also dimensionless, it can be used to break the mass-redshift degeneracy of the gravitational waveforms. 
Considering the observation through DECIGO, based on Fisher analysis, we have found that binary systems containing WDs are capable of breaking the mass-redshift degeneracy, achieving a redshift uncertainty $\Delta z/z < 0.1$. As the mass and frequency increase, the redshift uncertainty further decreases. Specifically at redshift $z>10^{-2}$, at low MT rates, the redshift uncertainty increases with rising redshift; in contrast, at high MT rates, the redshift uncertainty decreases as the redshift increases. Once the redshift is measured with high precision, we can utilize the redshift information provided by GWs to verify the accuracy of various cosmological models.

For binary systems in quasi circular orbits, the variation of the orbital angular frequency in Eq. \eqref{variationoemga} and the MT rate in Eq. \eqref{Chi*} exhibits high applicability when the WD mass ranges from 0.1 to 0.9 $M_\odot$. From evolutionary timescales of several years \cite{10.1093/mnras/stab626} to the 1 Gyr simulated by a Modules for Experiments in Stellar Astrophysics code (MESA) \cite{10.1093/mnrasl/slx064,Chen:2023etu,Chen:2022xhn,Chen_2020}, an agreement between the analytical and numerical solutions can be achieved by fine tuning the values of parameters $\mathcal{A}$ and $\mathcal{B}$ in Eq.\eqref{Chi*}, this significantly reduces the computational cost associated with MT binary systems.
When the MT rate is high enough $(\gtrsim 10^{-4} M_\odot/\text{yr})$, phenomena such as mass loss and accretion disk formation are expected to occur in the binary system. These effects have been neglected in the present study, and their incorporation is left for future work.
For higher MT rates, additional calculations are required to account for the dynamical effects of the accretion disk and common envelope.

In future work, we will further investigate the correlation between coefficients $\mathcal{A}$ and $\mathcal{B}$ in the MT rate in Eq.\eqref{Chi*} and the equation of state of WDs. In this work, we choose $\mathcal{A}=5\times 10^{-8}$ and $\mathcal{B}=7.8\times 10^{-5} M_\odot^{-4}$ to match the result in reference\cite{10.1093/mnras/stab626}. If we choose $\mathcal{A}'=10\mathcal{A}=5\times 10^{-7}$, there exists a relative variation of $2.27\%\thicksim 37.14\%$ in the maximum GW frequency, while the corresponding MT rate exists a relative variation of $2.86\%\thicksim 70.40\%$, with WD masses from $0.1 M_\odot$ to $0.9 M_\odot$. If we choose $\mathcal{B}'=0.5\mathcal{B}= 4\times 10^{-5}M_\odot^{-4}$, there exists a relative variation of $63.88\%\thicksim 75.29\%$ in the maximum GW frequency, while the corresponding MT rate exists a relative variation of $93.55\%\thicksim 97.59\%$. It can be found that the sensitivity of Eq.\eqref{Chi*} to parameter $\mathcal{A}$ is higher than that to $\mathcal{B}$, and the influence of parameters variations are smaller when the WD mass is low. $\mathcal{B}$ depends on the mean molecular weight per electron $\mu_e$ of the WD. This aims to derive a more precise analytical solution for the MT rate, enabling its application to a broader range of WD binary systems. 

\section{ACKNOWLEDGMENTS}
T. L. is supported by the China Postdoctoral Science Foundation Grant No. 2024M760692. This work is supported in part by the National Key Research and Development Program of China Grant No. 2020YFC2201501, in part by the National Natural Science Foundation of China under Grant No. 12475067 and No. 12235019, in part by the National Key Research and Development Program of China (No. 2021YFC2203003).

\appendix

\section{The variation of Mass transfer}\label{App.A}

To analyze the contributions of $\chi,\dot\chi$, and $\ddot\chi$ to the evolution of the frequency in Eq.\eqref{variationoemga} and masses of binary stars in Eq.\eqref{Deltat} and Eq.\eqref{Etat}, we come back to calculations in our previous work \cite{qhf1-v5jr}. We reconsider the MT variation terms and the quadratic MT terms. After order-of-magnitude estimation, for a MT rate of $10^{-3} M_\odot/\text{yr}$ at $0.1$ Hz, MT terms and MT variation terms make comparable contributions to the number of GW cycles, while quadratic MT terms are small quantities. Based on Eq.(B1) in \cite{qhf1-v5jr}, the complete 1PN Lagrangian in the center of mass frame that includes all mass transfer effects is expressed as
\begin{align}\label{Lagrangian}
    \mathcal{L}&=\frac{1}{2}M^2\eta v^2+\frac{G M^2\eta}{r}\notag\\
    &+\frac{1}{c^2}\bigg\{-\frac{G^2 M^3 \eta }{2 r^2}+\frac{G M^2\eta }{r} \left(\frac{1}{2} \eta v^2+\frac{3}{2} v^2+\frac{1}{2}\eta \dot{r}^2\right)-\frac{3}{8} M \eta ^2 v^4+\frac{1}{8} M \eta  v^4-\frac{1}{4} G M \dot{r} \Delta \chi -\frac{1}{4} G M r \Delta \dot\chi\bigg\},
\end{align}
where $c$ is the speed of light, $G$ is the gravitational constant, $v=\omega r$ is the velocity of binary, and $r$ is the radius of the orbit of the inspiral binary.

In the calculation process of the Euler-Lagrange equations, the MT variation terms are naturally canceled out, and there exist quadratic MT terms. We need to compare the contribution of $\chi,\chi^2$ and $\dot\chi$ in this equation as Table \ref{CompareA}.

\begin{table}[htbp!]
\small
  \centering
  \caption{The order of magnitudes of each MT terms in Eq.\eqref{variationoemga} with a dimension of $s^{-2}$. The initial conditions are $M=2 M_\odot$, $\omega=0.1\pi~\text{Hz}$, $\chi=10^{-3} M_\odot/\text{yr}$, and $\dot\chi=2\times10^{-4}M_\odot/\text{yr}^2$.}
    \renewcommand\arraystretch{1.6}
    \setlength{\tabcolsep}{10pt}
    \begin{tabular}{cccc}
    \hline
     0PN MT & 1PN MT & 1PN MT variation& 1PN quadric MT \\
    \hline  $\Delta\chi(M\eta)^{-1}$&$(GM)^{2/3}\omega^{2/3}\Delta\chi(M\eta)^{-1}$&$(GM)^{2/3}\omega^{5/3}\Delta\dot\chi(M\eta)^{-1}$&$(GM)^{2/3}\omega^{5/3}\Delta\chi^2(M\eta)^{-2}$\\
     $3.14\times 10^{-11}$&$6.74\times 10^{-15}$   & $1.36\times 10^{-22}$ & $2.16\times10^{-24}$ \\
    \hline
    \end{tabular}%
  \label{CompareA}%
\end{table}%
Table.\ref{CompareA} shows that, $\chi(0PN)\thicksim 10^4\times\chi(1PN)\thicksim 10^{11}\times\dot\chi(1PN)\thicksim 10^{13}\times\chi^2(1PN)$, these quadratic MT terms and MT variation terms at the 1PN order can be neglected as small quantities in the following calculations. For the Taylor expansion of WD mass $m_{WD}=m_{WD}(t=0)+\chi t+\dot\chi t^2/2+\ddot\chi t^3/6+\mathcal{O}(t^4)$, we estimate the order of magnitudes of above terms, MT rate $\chi,\dot\chi$ and $\ddot\chi$ product with observation time $t$, with a dimension of $M_\odot$

\begin{table}[htbp!]
\small
  \centering
  \caption{Taylor expansion of WD mass with initial condition $M_{WD}=0.8 M_\odot$, $\chi=10^{-3} M_\odot/\text{yr}$,$\dot\chi=2\times10^{-4} M_\odot/\text{yr}^2$, and observation time $t=5$yr. }
    \renewcommand\arraystretch{1.6}
    \setlength{\tabcolsep}{10pt}
    \begin{tabular}{cccc}
    \hline
     $m_{WD}$ & $\chi t$ & $\dot\chi t^2$& $\chi^2 t^2$ \\
    \hline  0.8&0.032&0.016&$9.95\times 10^{-4}$\\
    \hline
    \end{tabular}%
  \label{CompareB}%
\end{table}%
In Table.\ref{CompareB}, $(\chi/M) t \ll 1$ can be taken as a factor of Tyler expansion and contributions of $\dot\chi$ and $\chi^2$ can not be neglected to $M_{WD}$ and $\chi$.

Based on the above analysis of the variation of the MT rate, we get the following conclusion: the $\dot\chi$ appearing alone in the Lagrangian in Eq.\eqref{Lagrangian} and the evolution of frequency in Eq.\eqref{omega} can be neglected as a small quantity $\mathcal{O}(\chi^2)$ and $\mathcal{O}(\dot\chi)$, while the $\dot\chi t$ and $\ddot\chi t^2$ appearing in the mass cannot be neglected when the MT rate is about $\chi=10^{-3} M_\odot/\text{yr}$. This conclusion will be widely applied in the calculations of the main text.

\bibliography{BIB.bib}
\bibliographystyle{apsrev4-1}

\end{document}